\documentstyle[aas2pp4,psfig]{article}

\def\psr{PSR~B2303+46}
\def\MC{\ifmmode M_{\rm C}\else$M_{\rm C}$\fi}
\def\RC{\ifmmode R_{\rm C}\else$R_{\rm C}$\fi}
\def\taupsr{\ifmmode \tau_{\rm PSR}\else$\tau_{\rm PSR}$\fi}
\def\tpsr{\ifmmode t_{\rm PSR}\else$t_{\rm PSR}$\fi}
\def\tkin{\ifmmode t_{\rm kin}\else$t_{\rm kin}$\fi}
\def\tcool{\ifmmode t_{\rm cool}\else$t_{\rm cool}$\fi}
\def\tsn{\ifmmode t_{\rm SN}\else$t_{\rm SN}$\fi}
\def\Mpsr{\ifmmode M_{\rm PSR}\else$M_{\rm PSR}$\fi}
\def\Mcrit{\ifmmode M_{\rm crit}\else$M_{\rm crit}$\fi}
\def\Msun{\hbox{$M_\odot$}}
\def\Rsun{\hbox{$R_\odot$}}
\def\Teff{\ifmmode T_{\rm eff}\else$T_{\rm eff}$\fi}
\def\Mbol{\ifmmode M_{\rm bol}\else$M_{\rm bol}$\fi}
\def\un#1#2{\,\ifmmode{\rm #1^{#2}}\else$\rm #1^{#2}$\fi} 

\let\simgt\gtrsim\let\simlt\lesssim
\def\Sref#1{\S\ref{#1}}
\def\phnn{\phn\phn}

\makeatletter
\renewenvironment{deluxetable}[1]{\def\pt@format{\string#1}%
\set@tblnotetext\global\pt@ncol=0\global\pt@column=0\global\pt@page=1%
\def\pt@addcol{\global\advance\pt@ncol by\@ne}}%
{\pt@width\wd\pt@box\box\pt@box\vskip-0.5cm\spew@ptblnotes%
\typeout{Page \the\pt@page\space of table \thetable\space has been set to
width \the\pt@width\space with \the\pt@nlines\space lines per page}%
\endcenter\vskip-0.5cm\end@float}
\def\startdata{\pt@line=0\pt@calcnlines%
\ifdim\pt@width>\z@\def\@halignto{to \pt@width}\else\def\@halignto{}\fi%
\let\fnum@table=\fnum@ptable\set@mkcaption%
\@float{table}[t]\center\caption{\pt@caption}\leavevmode%
\setbox\pt@box=\pt@tabular{\pt@format}\pt@head}
\def\thebibliography{\subsection*{REFERENCES}
\list{}{\labelwidth3em\leftmargin\labelwidth\labelsep\z@\parsep\z@
\itemsep\z@\itemindent-3em\usecounter{enumi}}
\def\refpar{\relax}
\def\newblock{\hskip .11em plus .33em minus .07em}
\sloppy\clubpenalty4000\widowpenalty4000
\sfcode`\.=1000\relax}
\def\revtex@pageid{}
\makeatother

\begin{document}

\title{A Massive White Dwarf Companion to the Eccentric Binary Pulsar
System \psr} 

\righthead{White Dwarf Companion to \psr}

\author{M. H. van Kerkwijk}
\affil{Astronomical Institute, Utrecht University, P. O. Box 80000,
3508 TA Utrecht, The Netherlands}
\authoremail{M.H.vanKerkwijk@astro.uu.nl}
\and
\author{S. R. Kulkarni}
\affil{Palomar Observatory 105-24, California Institute of Technology, 
Pasadena, CA 91125, USA}
\authoremail{srk@astro.caltech.edu}

\begin{abstract}
Pulsars in close, eccentric binary systems are usually assumed to have
another neutron star as a companion.  These double neutron star
binaries have proven to be the best laboratories for experimental
General Relativity and are the most secure candidates for
gravitational wave interferometers.  We present deep $B$, $V$, and $R$
images of the field containing the eccentric binary pulsar system
\psr.  We find a faint, blue object ($B=26.60\pm0.09$;
$(B-R)_0=-0.4\pm0.2$) coincident with the timing position.  We suggest
this object is the optical counterpart to the \psr\ system.  The
counterpart is too bright to reflect emission from the pulsar or a
neutron star companion.  Most likely, the companion of \psr\ is not a
neutron star but a massive white dwarf.  We show that the observations
are consistent with a hot white dwarf companion
($\Teff\simgt5\times10^4$\un{K}{}) with cooling age equal the
characteristic age of the pulsar ($\tcool\simeq30\un{Myr}{}$) and mass
within the range set by timing observations and the Chandrasekhar mass
($1.2<\MC<1.4\un{\Msun}{}$).  Given the eccentric orbit, the white
dwarf must have formed before the neutron star, from what was
originally the more massive star in the binary.  Due to mass transfer,
the originally less massive star could become sufficiently massive to
end its life in a supernova explosion and form the radio pulsar.  We
constrain the mass of the pulsar to be in the range
$1.24<\Mpsr<1.44\un{\Msun}{}$.
\end{abstract}


\section{Introduction}\label{sec:intro}

Binaries in which a radio pulsar is in a close, eccentric orbit around
a compact companion have provided the best astronomical tests of
General Relativity and the most accurate neutron-star mass
determinations (Taylor et al.\ \cite{tayl&a:92}); are the most secure
targets for gravitational wave interferometers such as LIGO
(Abramovici et al.\ \cite{abra&a:92}); and may be the progenitors of
the enigmatic gamma-ray bursts (Piran \cite{pira:97}).  Especially for
the latter two topics, the nature of the compact companion is of vital
importance.  Fortunately, optical observations can distinguish the
three possibilities: white dwarfs can be detected out to large
distances, whereas neutron stars are rather dim and black holes not
detectable at all.

It has become general practice to identify as double neutron star
system any high-eccentricity binary pulsar for which the inferred
companion mass is $\sim\!1.4\un{\Msun}{}$ (for a review, Van den
Heuvel \cite{vdhe:95}).  This is based on the evolutionary scenario
for these systems.  Briefly, it starts with an early-type binary.  The
primary (the more massive star) evolves first, transfers some mass to
the secondary, explodes as a supernova, and forms a neutron star.
Next, the secondary evolves and starts to transfer matter to the
neutron star.  The accretion causes ``recycling'': spin-up and -- in a
way not well understood -- a reduction in magnetic field strength.
The mass transfer will be unstable, leading to a common-envelope
phase.  The orbit will necessarily be circularized; it can only become
eccentric if the secondary is sufficiently massive to explode and form
a neutron star in turn.  (Otherwise, a white dwarf is left in a
circular orbit, as observed for other binary pulsars.)  Thus, if the
observed radio pulsar can be shown to be the first-formed, recycled
neutron star, the companion must be a neutron star as well.

In four of the six presumed double-neutron star systems --
PSR~B1913+16, J1518+4904, B1534+12, and B2127+11C -- the pulsars
indeed appear recycled: the spin periods are shorter and inferred
magnetic fields weaker than for ordinary pulsars (tens of ms vs.\
$\sim\!1\un{s}{}$, and $\sim\!10^{10}\un{G}{}$ vs.\
$\simgt\!10^{12}\un{G}{}$).  However, the pulsars in the other two
systems, PSR~B1820$-$11 and B2303+46, show no clear sign of recycling.
For the former, Phinney \& Verbunt (\cite{phinv:91}) suggested that
the companion was not a neutron star but a low-mass main-sequence
star, and that the system will eventually become a low-mass X-ray
binary.  The latter, \psr, is the subject of this paper.

\section{PSR~B2303+46}\label{sec:psr}

\psr\ is in a 12.3\un{d}{}, highly eccentric ($e=0.66$) orbit (Stokes,
Taylor, \& Dewey \cite{stoktd:85}).  Periastron advance is observed,
from which one infers a total mass of the system
$\Mpsr+\MC=2.64\pm0.05\un{\Msun}{}$, as well as, in combination with
the mass function, the limits $\Mpsr<1.44\un{\Msun}{}$ and
$\MC>1.20\un{\Msun}{}$ (Thorsett et al.\ \cite{thor&a:93}; Thorsett \&
Chakrabarty \cite{thorc:98}).

The pulsar period ($P=1.06\un{s}{}$) and inferred magnetic field
strength ($B\simeq8\times10^{11}\un{G}{}$) allow the possibility of
mild recycling.  Given the small characteristic age,
$\taupsr=P/2\dot{P}\simeq30\un{Myr}{}$, radio pulsations of the
presumed neutron-star companion have been searched for, but with no
success.  Kulkarni (\cite{kulk:88}) carried out optical observations
with the Palomar 200-inch, but found no counterpart down to
$R=26\un{mag}{}$; this was seen as confirmation of the scenario
outlined above.

The pulsar parameters, however, are also consistent with those of
ordinary pulsars.  Thus, the pulsar could have have formed after the
companion completed its evolution.  If so, it may be the only neutron
star in the binary, the companion being a white dwarf.  This requires
a twist to the evolutionary scenario, in which one starts with two
stars with masses (slightly) below the critical mass, \Mcrit
($\sim\!8\un{\Msun}{}$; Koester \& Reimers \cite{koesr:96}), required
to evolve to a neutron star.  In due course, the primary evolves,
transfers matter to the secondary, and forms a white dwarf.  Now if
the mass transfer increased the secondary mass beyond \Mcrit, it can
explode and form a neutron star, resulting in a binary with an older
white dwarf and a younger neutron star in a highly eccentric orbit.

We were reminded of this possibility by Dr Wij\-ers (1997, private
communication), who wondered whether it could be verified
observationally.  At a distance of 4.3\,kpc (inferred from the pulsar
dispersion measure) and for a cooling age of 30\un{Myr}{} (\taupsr), a
$1.2\un{\Msun}{}$ white dwarf counterpart would have $V\simeq25$.
This is excluded by the limit mentioned above, but, perhaps
fortuitously, we had forgotten about this result.  Here, we report
new, deeper optical observations with the Keck telescope, which show a
possible counterpart to the \psr\ system.

\section{Optical Observations}\label{sec:obs}

We imaged the field containing \psr\ with the Low Resolution Imaging
Spectrograph (Oke et al.\ \cite{oke&a:95}) at the Keck~II telescope,
on the nights of November 28 and 29, 1997 (UT).  On the 28th, three
600-s exposures were obtained in the R band, and two 900-s exposures
in~B.  On the 29th, one 450-s and four 600-s in~R, three 900-s in~B,
and five 600-s exposures in~V were taken.  All images were taken at
airmass $<\!1.4$.  The skies were clear on the second night, but the
first night was plagued by cirrus.

The reduction was done as described by Kulkarni \& van Kerkwijk
(\cite{kulkvk:98}) for the field of RX~J0720.4$-$3125, which was
observed on the same nights.  For the photometric calibration, we used
Landolt fields: in B and R, the four listed in Table~1 of the above
reference; in V, the first two only.  We estimate uncertainties of
$\simlt\!0.02\un{mag}{}$ in the zero points.

For the astrometry, we selected from the USNO-A2.0 catalogue (Monet et
al.\ \cite{mone&a:98}) all 163 stars that overlapped with a 10-s
R-band image.  We measured their centroids, and corrected for
instrumental distortion using a bi-cubic function determined by Cohen
(private communication).  With the plate scale known accurately, we
fitted only for the zero-points in each coordinate and the position
angle on the sky.  After rejecting 7 outliers (residual larger than
0\farcs8), the root-mean-square residuals were 0\farcs20 in each
coordinate.  The astrometry was transferred to the stacked $B$, $V$,
and $R$ images using 28 transfer stars close to the pulsar position,
solving again for rotation and zero-points.  The rms residuals were
$\simlt\!0\farcs04$.

\begin{figure*}
\centerline{\hbox{\psfig{figure=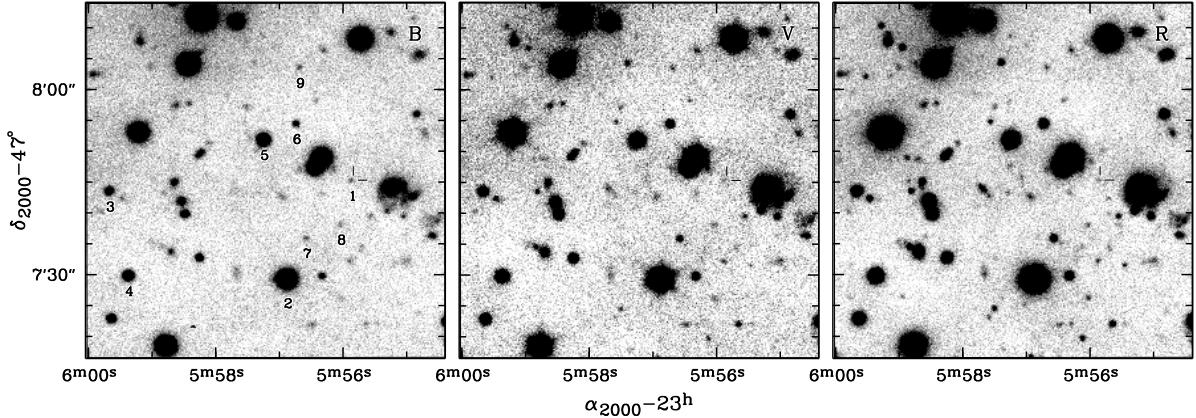,width=0.9\textwidth,angle=-90}}}
\caption[]{Stacked B, V, and R images of the localization of \psr.  In
each, the timing position is indicated with the tick marks (of size
1\farcs5).  Stars mentioned in the text are labeled below their
image.\label{fig:images}}
\end{figure*}

Close to the timing position of \psr, we found one faint, relatively
blue object, hereafter star~1; see Figure~\ref{fig:images} and
Table~\ref{tab:phot-astr}.  To verify whether the respective
positions are consistent with each other, one has to take into account
the measurement uncertainties ($\sigma_1=0\farcs04$,
$\sigma_{\rm{}PSR}=0\farcs17$ in each coordinate), as well as the
extent to which the two positions are on the same astrometric system.
The USNO-A2.0 catalogue is tied to the International Celestial
Reference System as realized by the Tycho-based ACT catalogue (see
Monet et al.\ \cite{mone&a:98}), while the DE200 dynamical ephemeris
-- on which the pulsar timing position is based -- is close to the
ICRS as well (Folkner et al.\ \cite{folk&a:94}).  We expect the frame
difference between DE200 and USNO-A2.0 to be considerably less than
$\sigma_{\rm{}PSR}$.  Star~1 is offset by 0\farcs28 from the timing
position, i.e., well within the 95\% confidence radius of 0\farcs4 one
infers from $\sigma_{\rm{}PSR}$ alone.

\begin{deluxetable}{lccccc}
\tabcolsep0in\footnotesize
\tablewidth{\hsize}
\tablecaption{Photometry and Astrometry\label{tab:phot-astr}}
\tablehead{%
\colhead{Star}&
\colhead{$\alpha_{\rm J2000}$}& 
\colhead{$\delta_{\rm J2000}$}&
\colhead{B}&
\colhead{V}&
\colhead{R}\nl
\colhead{}&
\colhead{\rlap{(}\phnn\rlap{$^{\rm h}$}~\phnn
\rlap{$^{\rm m}$}~\phnn\rlap{$^{\rm s}$}\phd\phnn\phn\llap{)}}&
\colhead{\rlap{(}\phnn\rlap{\arcdeg}~\phnn
\rlap{\arcmin}~\phnn\rlap{\arcsec}\phd\phnn\llap{)}}&
\colhead{(mag)}&
\colhead{(mag)}&
\colhead{(mag)}}
\startdata
PSR&23 05 55.842&47 07 45.32\nl
  1&23 05 55.869&47 07 45.30&26.60(\phn9)&26.91(20)&   26.65(16)\nl
  2&23 05 56.875&47 07 29.37&19.84(\phn2)&18.89(\phn2)&\nodata\nl
  3&23 05 59.665&47 07 43.54&24.08(\phn2)&22.64(\phn2)&21.72(\phn2)\nl
  4&23 05 59.362&47 07 29.88&22.79(\phn2)&21.16(\phn2)&20.10(\phn2)\nl
  5&23 05 57.242&47 07 51.89&22.00(\phn2)&20.62(\phn2)&19.69(\phn2)\nl
  6&23 05 56.734&47 07 54.47&25.18(\phn3)&23.46(\phn2)&22.21(\phn2)\nl
  7&23 05 56.574&47 07 35.96&26.51(\phn8)&24.74(\phn3)&23.40(\phn2)\nl
  8&23 05 56.036&47 07 38.18&26.73(10)&   26.39(11)&   25.99(\phn9)\nl
  9&23 05 56.682&47 08 03.55&26.43(\phn8)&26.20(\phn9)&25.87(\phn8)\nl
\enddata
\tablecomments{The top line gives the timing position for \psr\
(Thorsett et al.\ \cite{thor&a:93}); the uncertainties are 0\fs017 and
0\farcs17.  Star~1 is the proposed optical counterpart; other stars
are discussed in the text.  The measurement uncertainties for the
positions of optical objects are $\simlt\!0\fs004$ and
$\simlt\!0\farcs04$.  Possible uncertainties in the tie to astrometric
systems are discussed in \Sref{sec:obs}.  The uncertainties in the
photometry are indicated by the numbers in brackets.  Star~2 has no
R-band magnitude, as it was overexposed in the R-band images.}
\end{deluxetable}

For the photometry, we used a simple point-spread function fitting
method, which takes into account gradients in the sky level (see
Kulkarni \& van Kerkwijk \cite{kulkvk:98}).  First, we used aperture
photometry on images from the 29th to measure the instrumental
magnitudes for relatively isolated, brighter ``secondary'' stars
(stars 2--6 in B, V; 3--6 in R, because star 2 was overexposed;
Fig.~\ref{fig:images} and Table~\ref{tab:phot-astr}).  Next, we
extracted from the stacked images $21\times21$ pixel
($4\farcs5\times4\farcs5$) regions around stars 2--6, the candidate
(star~1), and three other faint objects (stars 7--9).  We fitted these
to a 2-dimensional Gaussian on top of a plane with an arbitrary tilt,
and determined the average FWHM for the secondary stars.  We then
refitted all objects keeping the FWHM fixed at the average, and used
the amplitudes of the Gaussians to determine relative magnitudes.
Finally, the difference with the aperture results for the secondary
stars was used to calculate instrumental magnitudes for stars 1, 7, 8,
and~9, and all magnitudes were calibrated using the solution found
from the Landolt stars.  The results are listed in
Table~\ref{tab:phot-astr}.  In order to verify our procedures, we also
determined $B$ and $R$ magnitudes for stars 1, 7, 8, and~9 from the
stacked images from the 28th and 29th separately: these gave
consistent results.

\section{A Massive White Dwarf Companion}\label{sec:companion}

The probability that star~1 is a background object that happens to be
within the 0.5\un{arcsec}{2} error region (95\% confidence) of the
timing position of \psr\ is about 5\%, i.e., not particularly low.
With $B-R=-0.05\pm0.18$, however, star~1 is bluer than all other faint
stars in the field.  In our images, the bluest other objects have
$B-R\simeq0.6$ (e.g., stars 8 and~9).  Of these, there are only a few
arcmin$^{-2}$ and the chance coincidence probability is $<\!0.1\%$.
The probability for an object as blue as star~1 is lower still, and,
therefore, we believe star~1 is the optical counterpart to \psr.

It is unlikely that the optical emission is due to the pulsar or to a
neutron-star companion.  Thermal emission from a neutron star could
reproduce the colors, but it would be much too faint: the known
sources have similar magnitudes, but are all nearby (for a recent
compilation, see Mignani \cite{mign:98}).  Nonthermal emission can
lead to brighter sources, but only for young pulsars and generally
with colors that are too red.  This leads us to propose that the
companion of \psr\ is a massive white dwarf, and that star~1 is its
optical counterpart.

In order to verify whether our observations are consistent with a
white-dwarf companion, we need to estimate the expected brightness.
This is possible using cooling models for white dwarfs, provided we
have estimates for the white-dwarf mass, composition, and age, as well
as for the distance and reddening.  We will discuss these in turn.

The mass of the white dwarf companion has a strict lower bound of
1.2\un{\Msun}{} (inferred from timing; \Sref{sec:psr}).  An equally
strict upper bound of 1.4\un{\Msun}{} is set by the Chandrasekhar
mass.

The age of the white dwarf is the sum of \tsn,
the time that elapsed between the formation of the white dwarf and the
supernova explosion, and \tpsr, the age of the pulsar (see
\Sref{sec:psr}).  An upper limit to \tsn\ is set by the total lifetime
of an 8-\Msun (\Mcrit) star, i.e., $\tsn<40\un{Myr}{}$ (e.g., Schaller
et al.\ \cite{scha&a:96}).  An upper limit to \tpsr\ is set by the
characteristic age (but see \Sref{sec:ramif}), i.e.,
$\tpsr\simlt\taupsr=30\un{Myr}{}$.

The cooling age of the white dwarf may equal its actual age, or
$\tcool=\tsn+\tpsr<70\un{Myr}{}$.  It is quite likely, however, that
the white dwarf was reheated when, prior to the supernova explosion,
the system went through a common-envelope phase (required to account
for the current small orbital size).  If so, $t_{\rm{}SN}$ is
irrelevant, and $\tcool\simeq\tpsr\simlt30\un{Myr}{}$.

The composition of massive white dwarfs is still uncertain,
but observations of novae indicate that both C+O and O+Ne+Mg are
possible\footnote{For some massive white dwarfs, masses and radii
indicate Fe composition (Provencal et al. \cite{prov&a:98}).  This is
excluded here: for Fe white dwarfs $M_{\rm{}max}=1.1\un{\Msun}{}$
(Hamada \&\ Salpeter \cite{hamas:61}).} (Starrfield \cite{star:89}).
Fortunately, the mass-radius relations are very similar (as inferred
from Hamada \& Salpeter \cite{hamas:61}), and hence so should the
cooling tracks.  For massive white dwarfs, effects related to the
composition of the atmosphere are expected to be small as well (Wood
\cite{wood:95}).  For completeness, we note that a hydrogen atmosphere
seems likely, as some hydrogen will have been accreted during the
common-envelope phase.

The distance towards \psr\ can be constrained from the observed
dispersion measure (DM) of $62\un{cm}{-3}\un{pc}{}$. Using the
Galactic model for the distribution of free electrons of Taylor \&
Cordes (\cite{taylc:93}), we find that in the pulsar's direction
($l^{\rm{}II}=105\fdg41$, $b^{\rm{}II}=-11\fdg93$) the predicted DM is
consistent with the observed one for any distance $d>2.5\un{kpc}{}$.
There is no upper limit on the distance, as the predicted maximum DM
(for objects well outside the electron layer) is
$78\pm20\un{cm}{-3}\un{pc}{}$.  At the lower limit, the predicted DM
is $41\pm10\un{cm}{-3}\un{pc}{}$.

The reddening along the line of sight, estimated from dust infrared
emission, is $E_{B-V}=0.22\pm0.03$ (Schlegel, Finkbeiner, \& Davis
\cite{schlfd:98}).  One infers $A_B=0.95\pm0.13$, $A_V=0.73\pm0.10$,
and $A_R=0.59\pm0.08$.  For any likely identification, star~1 is well
out of the Galactic plane, and the full reddening should be taken into
account.

With the above, we are in a position to estimate the cooling flux from
a presumed white dwarf companion to \psr.  From the cooling tracks of
Benvenuto \& Althaus (\cite{benva:99}), we find that after
30\un{Myr}{} a 1.2\un{\Msun}{} C+O white dwarf has cooled down to
$\Mbol=6.5$ and $\Teff=5\times10^4\un{K}{}$.  (Similar results are
obtained extrapolating tracks of Wood [\cite{wood:95}]).  From the
atmospheric models of Bergeron, Wesemael, \& Beauchamp
(\cite{bergwb:95}), we find for this temperature $BC_V=-4.5$ and
Rayleigh-Jeans-like colors $(B-V)_0=-0.28$, $(V-R)_0=-0.14$ (for a
hydrogen atmosphere; differences for other atmospheric compositions
should be small).  Thus, one expects $M_B=10.7$ and $(B-R)_0=-0.42$.

For a cooling age of 70\un{Myr}{}, we find $\Mbol=7.5$,
$\Teff=4\times10^4\un{K}{}$, $M_B=11.0$, $(B-R)_0=-0.42$.  For higher
masses, predictions are harder to make, as no cooling tracks are
available.  Extrapolating, we expect more massive white dwarfs to have
smaller radii, but be hotter at the same age.  The likely net effect
will be that they are brighter bolometrically, but fainter in the
optical (showing the same color).  A conservative lower limit
$M_B<13.4$ is inferred by taking into account the change in radius
only, from $\sim\!0.006\un{\Rsun}{}$ for a 1.2\un{\Msun}{} white dwarf
to $\sim\!0.002\un{\Rsun}{}$ for one at the Chandrasekhar mass (Hamada
\& Salpeter \cite{hamas:61}).

In summary, we expect that a white-dwarf companion will have
$(B-R)_0=-0.42$ and $13.4>M_B\simgt10.7$.  It can be brighter only if
$\tcool$ is substantially smaller than $\taupsr$, which we consider
unlikely.  The expected reddened color, $B-R=-0.06\pm0.05$, is
consistent with the observations.  The expected distance modulus is
$12.3<B-M_B-A_B\simlt15$, corresponding to a distance of
$3<d\simlt\!10\un{kpc}{}$, consistent with the lower limit set by the
dispersion measure.

\section{Ramifications}\label{sec:ramif}

We presented Keck imaging of \psr, in which we identified a faint
blue object, star~1, coincident with the precise timing position.  We
have shown that the companion to \psr\ could be a hot, massive white
dwarf, with star~1 its optical counterpart.  We have outlined how the
evolution of a binary composed of two stars with masses close to but
below \Mcrit\ could lead to the formation of a system like \psr.  

The system may help calibrate different chronometers.  With a
temperature measurement from UV observations ({\em{}HST} time
granted), cooling models will provide a lower limit to \tcool.  This
may allow a test of pulsar braking: if one finds
$\tcool\gg\tsn+\taupsr$, the braking index $n$ has to be the culprit.
The braking index enters via $\taupsr=P/(n-1)\dot{P}$.  Usually, $n=3$
is assumed (as we have done above), which is valid for a dipole
rotating in vacuo.  However, this has not been verified
observationally.  Indeed, all measurements give $n<3$, although these
were for some young, fast pulsars, for which $n<3$ is perhaps expected
(Melatos \cite{mela:97}).  It would be interesting as well to find
$\tcool\ll\taupsr$: this would imply that pulsars do not have to be
born with short spin periods.


If the companion is indeed a white dwarf, both the lower and the upper
limit to its mass are interesting.  The upper limit, in combination
with the total mass, corresponds to a lower limit to the mass of the
pulsar.  Combined with the upper limit set by timing (\Sref{sec:psr}),
one finds $1.24<\Mpsr<1.44\un{\Msun}{}$.

The lower limit of 1.2\un{\Msun}{} makes the companion interesting as
a white dwarf.  Masses $>\!1\un{\Msun}{}$ have also been inferred for
about a dozen field white dwarfs, many discovered only recently from
extreme UV sky surveys (Marsh et al.\ \cite{mars&a:97}; Vennes et al.\
\cite{venn&a:97}; Finley, Koester, \& Basri \cite{finlkb:97}).  It is
not clear, however, whether these have been formed from massive stars,
for the following reasons.  First, quite a few, especially the more
massive ones, have strong magnetic fields, which is unusual; also, it
makes the mass estimates, which are based on line broadening, more
uncertain (e.g., Schmidt et al.\ \cite{schm&a:92}; Ferrario, Vennes,
\& Wickramasinghe 1998).  Second, most are hot and therefore young;
combined with the short lifetimes of stars massive enough to form
them, one would expect to find them in young star clusters, not
in the field (Bergeron et al.\ \cite{berg&a:91}).  Third, in at least
one object, GD\,50, unexpected traces of helium have been found in the
spectrum, as well as evidence for a high rotation rate (Vennes,
Bowyer, \& Dupuis \cite{vennbd:96}).  Fourth, the massive white dwarfs
appear to form not just a tail of the distribution of white-dwarf
masses, but rather a separate peak (Finley et al.\ \cite{finlkb:97}).
These reasons have led to the speculation that at least some of these
field massive white dwarfs are not the product of single-star
evolution, but rather the result of mergers of two ordinary
0.6\un{\Msun}{} white dwarfs (references cited above).

White dwarfs in young star clusters and binaries almost certainly are
the product of massive stars, but these have masses up to
$\sim\!1\un{\Msun}{}$ only (in NGC~2516 [Koester \&\ Reimers
\cite{koesr:96}] and Sirius [Gatewood \& Gatewood \cite{gateg:78};
Provencal et al.\ \cite{prov&a:98}]).  In contrast, the massive white
dwarf companion of \psr\ has almost certainly descended from a massive
star and it is undeniably massive -- a statement that can be made
given the exquisite precision of pulsar timing.

\acknowledgments We are grateful to Dave Monet for providing us with
preliminary astrometric results from USNO-A2.0.  We acknowledge
support by a fellowship of the Royal Netherlands Academy of Arts and
Sciences (MHvK), and by grants from NASA and NSF (SRK).  The
observations reported here were obtained at the W. M. Keck Observatory
which is operated by the California Association for Research in
Astronomy, a scientific partnership among California Institute of
Technology, the University of California and the National Aeronautics
and Space Administration.  It was made possible by the generous
financial support of the W. M. Keck Foundation.

\end{document}